\documentclass[preprint2,trackchanges,lineno]{aastex6}
\usepackage{newtxtext,newtxmath,hyperref}
\usepackage{ae,aecompl}
\usepackage{graphicx}	% Including figure files
\usepackage{amsmath}	% Advanced maths commands
\usepackage{amssymb,color}	% Extra maths symbols

\newcommand{\Msun}{\ensuremath{\mathrm{M}_\odot}}

\slugcomment{draft v1}

\shorttitle{Delayed Jet Breakouts from Binary Neutron Star Mergers}
\shortauthors{Matsumoto \& Kimura}

\begin{document}

\title{Delayed Jet Breakouts from Binary Neutron Star Mergers}

\author{Tatsuya Matsumoto\altaffilmark{1,2,3}}
\email{tatsuya.matsumoto@mail.huji.ac.il}
\author{Shigeo S. Kimura\altaffilmark{4,5,6}}

\altaffiltext{1}{Racah Institute of Physics, Hebrew University, Jerusalem, 91904, Israel}
\altaffiltext{2}{Department of Physics, Graduate School of Science, Kyoto University, Kyoto 606-8502, Japan}
\altaffiltext{3}{JSPS Research Fellow}
\altaffiltext{4}{Department of Physics, The Pennsylvania State University, University Park, Pennsylvania 16802, USA}
\altaffiltext{5}{Department of Astronomy \& Astrophysics, The Pennsylvania State University, University Park, Pennsylvania 16802, USA}
\altaffiltext{6}{Center for Particle and Gravitational Astrophysics, The Pennsylvania State University, University Park, Pennsylvania 16802, USA}

\begin{abstract}
Short gamma-ray bursts (sGRBs) are thought to be produced by binary NS mergers.
While a sGRB requires a relativistic jet to break out of ejecta, the jet may be choked and fails to produce a successful sGRB.
We propose a ``delayed breakout'' scenario where a late-time jet launched by a long-term engine activity can penetrate ejecta even if a prompt jet is choked.
Observationally, such a late-time jet is supported by the long-lasting high-energy emissions in sGRBs.
Solving the jet propagation in ejecta, we show that a typical late-time activity easily achieves the delayed breakout.
This event shows not prompt $\gamma$-rays but long-time X-ray emissions for $\sim10^{2-3}{\,\rm s}$ or even $\sim10^{4-5}{\,\rm s}$.
Some delayed events may be already detected as soft-long GRBs without supernova signatures.
In an optimistic case, a few events coincident with gravitational-waves (GWs) are detected by the second-generation GW detectors every year. 
X-ray followups of merger events without $\gamma$-rays will be a probe of long-lasting engine activities in binary mergers.
%We also discuss an implication of a characteristic jet-breakout time in sGRBs.
\end{abstract}


\section{introduction}\label{introduction}
Short gamma-ray bursts (sGRBs) are a class of GRBs whose duration is less than $2{\,\rm s}$ \citep[][for reviews]{Nakar2007,Berger2014}.
They are believed to be powered by relativistic jets launched from compact binary mergers \citep{Eichler+1989}.
This model is strongly supported by the detection of the gravitational waves (GWs) from the merging binary neutron star (NS) \citep[GW170817,][]{Abbott+2017c} and by the extensive followups of electromagnetic counterparts, especially VLBI observations \citep{Mooley+2018b,Ghirlanda+2018}. 

Some binary NS mergers may fail to produce sGRBs even if they launch relativistic jets.
In order to produce a sGRB, the relativistic jet should break out of the matter ejected by the binary coalescence \citep{Nagakura+2014,Murguia-Berthier+2014}.
When the ejecta are too massive or the jet opening angle is too large, the jet is choked and fails to emit prompt $\gamma$-rays.
Choked-jet events are supported by a threshold timescale to produce sGRBs which appears in the duration distribution of sGRBs \citep{Moharana&Piran2017}.
Furthermore, the observed binary-NS-merger rate larger than the local sGRB rate also suggests choked events.

In addition to prompt $\gamma$-rays, some (or most) sGRBs show long-lasting high energy emissions \citep{Kisaka+2017}.
They are classified into an extended emission with the duration of $\sim10^{2-3}{\,\rm s}$ \citep{Norris&Bonnell2006} and a plateau emission with $\sim10^{4-5}{\,\rm s}$ \citep{Gompertz+2013,Gompertz+2014}.
Since it is difficult to explain them in the standard afterglow theory \citep{Ioka+2005}, their origins are attributed to prolonged central-engine activities which launch jets or outflows \citep{Metzger+2008,Nakamura+2014,Kisaka&Ioka2015}.

\begin{figure}
\begin{center}
\includegraphics[width=60mm, angle=90]{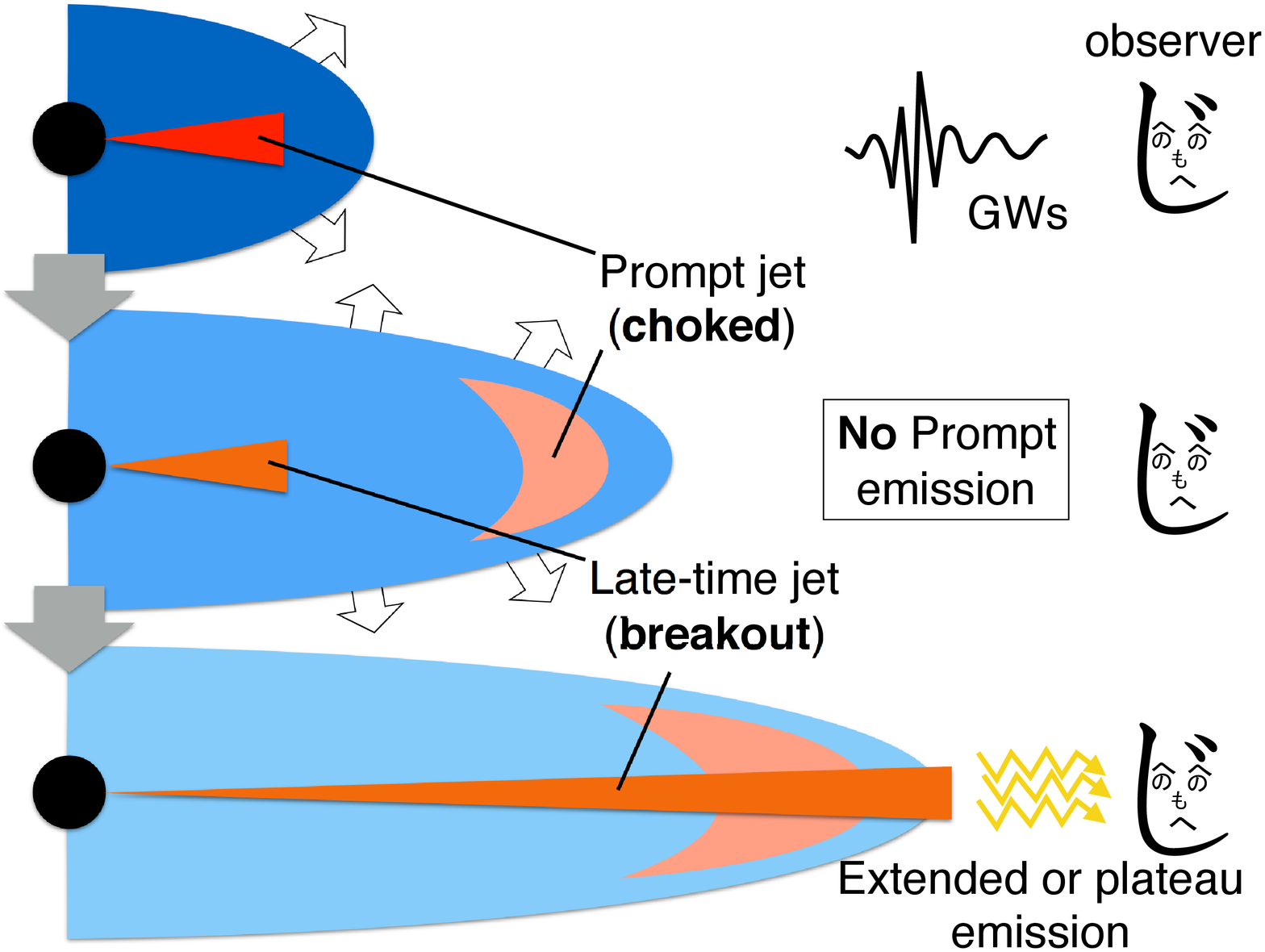}
\caption{
Schematic picture of a delayed-jet-breakout event.
First, a prompt jet is choked and fails to produce a sGRB (top).
Another jet which powers extended or plateau emission is launched later (middle).
It is also possible that the delayed jet is an identical jet as the prompt one but with reduced luminosity.
Due to expansion, the ejecta density becomes tenuous and helps the late-time jet to break out of the ejecta (bottom).
}
\label{fig picture}
\end{center}
\end{figure}

Even if a prompt jet is choked, a late jet may penetrate ejecta.\footnote{We do not specify whether a delayed jet is powered separately from a prompt jet \citep{Metzger+2008} or  the same as the prompt one but with reduced luminosity \citep{Kisaka&Ioka2015}. For the latter case, after the prompt jet is choked, a prolonged energy injection from the engine may produce a jet head structure.}
Late-time jets can be more powerful than prompt ones because some extended emissions have larger energy than that of prompt emissions \citep[e.g.,][]{Perley+2009}.
The ejecta's expansion also helps the late jet to break out by reducing the ejecta density.
Hereafter, we call this scenario as ``delayed jet breakout''.
In Fig. \ref{fig picture}, we show a schematic picture. 
We calculate the propagation of the late-time jet in the ejecta and find that the delayed jet breakout is realized with a typical late-time engine activity.
In such an event, we cannot detect prompt $\gamma$-rays because the prompt jet is choked. 
Instead a late-time jet breaks out of the ejecta $\gtrsim10^{1-2}{\,\rm s}$ after the merger, and produces extended and plateau emissions. This can be observed as a soft-long GRB.
We also discuss the event rate of the delayed jet breakouts and argue that they might have been observed by \textit{MAXI}.

\section{Jet Propagation in expanding media}\label{Jet Propagation}
We calculate the jet propagation in ejecta of a binary NS merger by using a semi-analytical formula along \cite{Bromberg+2011} \citep[see also][]{Margalit+2018}.
A jet launched from a central engine collides with ejecta and produces a jet head and cocoon.
The cocoon surrounds the jet and collimates it (see below).
The jet head velocity is given by the momentum balancing at the head as \citep{Matzner2003,Bromberg+2011}
\begin{eqnarray}
\beta_{\rm h}&=&\frac{{\tilde L}^{1/2}\beta_{\rm j}+\beta_{\rm a}}{{\tilde L}^{1/2}+1},
   \label{jet head}\\
{\tilde L}&\simeq&\frac{L_{\rm j}}{\Sigma_{\rm j}\rho_{\rm a}\Gamma_{\rm a}^2c^3},
   \label{L tilde}
\end{eqnarray}
where $\beta_{\rm j}\simeq1$, $\beta_{\rm a}$, $L_{\rm j}$, $\Sigma_{\rm j}$, $\rho_{\rm a}$, $\Gamma_{\rm a}=(1-\beta_{\rm a}^2)^{-1/2}$, and $c$ are the velocity of the jet material, ejecta velocity, one-sided jet luminosity,\footnote{This luminosity is written as $L_{\rm j}=\theta_{\rm j}^2L_{\rm j,iso}/4$ by the jet opening angle $\theta_{\rm j}$ and isotropic jet luminosity $L_{\rm j,iso}$.} jet cross section, ejecta density, ejecta's Lorentz factor, and speed of light, respectively.
The ejecta's quantities are evaluated at the head. 

We assume that the ejecta are homologous and have a power-law density profile \citep{Hotokezaka+2013,Nagakura+2014}.
Even after a prompt jet is choked, although the profile is modified from the original one, these assumptions may hold.
Note that if a prompt jet succeeds in breaking out, it produces a cavity and a late jet (or even a spherical outflow) easily emerges from the ejecta (see also \S \ref{As a probe of late-time engine activity in binary NS mergers}).
For homologous ejecta, the velocity is given by $\beta_{\rm a}=(R_{\rm h}/R_{\rm ej})\beta_{\rm ej}$, where $R_{\rm h}$, $R_{\rm ej}=\beta_{\rm ej}c(t+t_{\rm lag})$, and $\beta_{\rm ej}$ are the jet head position and the radius and velocity of the ejecta edge, respectively.
We set the origin of time $t$ as the jet-launching time, which is $t_{\rm{lag}}$ after the merger.
The density is given by
\begin{eqnarray}
\rho_{\rm a}&=&\frac{M_{\rm ej}}{4\pi{R_{\rm ej}^3}}f\biggl(\frac{R_{\rm h}}{R_{\rm ej}}\biggl)^{-k}\,\,\,{\rm for}\,\,\,R_{\rm in}\leq{R}\leq{R_{\rm ej}},\\
f&=&\begin{cases}
\frac{3-k}{1-\bigl(\frac{\beta_{\rm esc}}{\beta_{\rm ej}}\bigl)^{3-k}}&k\neq3,\\
\frac{1}{\ln(\beta_{\rm ej}/{\beta_{\rm esc}})}&k=3,
\end{cases}
\end{eqnarray}
where $M_{\rm ej}$ is the ejecta mass.
The inner boundary is set by the innermost unbound ejecta at the jet launch as
\begin{eqnarray}
\beta_{\rm esc}=\biggl(\frac{GM_{\rm c}\beta_{\rm ej}}{c^2R_{\rm ej}}\biggl)^{1/3}\simeq0.023\,\biggl(\frac{t_{\rm lag}}{\rm s}\biggl)^{-1/3}\biggl(\frac{M_{\rm{c}}}{2.6\,\Msun}\biggl)^{1/3},
\end{eqnarray}
where $G$ and $M_{\rm c}$ are the gravitational constant and the merger-remnant mass, respectively.

The cocoon pressure determines whether the jet is collimated or conical.
The jet cross section is given as
\begin{eqnarray}
\Sigma_{\rm j}=\begin{cases}
\pi\theta_{\rm j}^2R_{\rm h}^2&\text{conical jet},\\
\frac{L_{\rm j}\theta_{\rm j}^2}{4cP_{\rm c}}&\text{collimated jet}.
\end{cases}
\end{eqnarray}
The cocoon pressure is given by
\begin{eqnarray}
P_{\rm c}=\frac{E_{\rm c}}{3V_{\rm c}}=\frac{\int{L_{\rm j}(1-\beta_{\rm h})dt}}{\pi{}R_{\rm c}^2R_{\rm h}},
   \label{cocoon pressure}
\end{eqnarray}
where the cocoon is radiation-pressure dominated and conical with a height $R_{\rm h}$ and radius $R_{\rm c}$.
The cocoon radius is obtained by integrating the cocoon's lateral-expansion velocity of \citep{Begelman&Cioffi1989}
\begin{eqnarray}
\beta_{\rm a}=\sqrt{\frac{P_{\rm c}}{\bar{\rho}_{\rm a}c^2}},
   \label{cocoon velocity}
\end{eqnarray}
where $\bar{\rho}_{\rm a}$ is the cocoon's mean density.
When a converging position of the jet's collimation shock \citep{Komissarov&Falle1997}
\begin{eqnarray}
\hat{z}=\sqrt{\frac{L_{\rm j}}{\pi{cP_{\rm c}}}}
\end{eqnarray}
is lower than the jet head $R_{\rm h}\gtrsim{\hat{z}}$, the jet is collimated.

We integrate above equations numerically and obtain the jet-breakout time $t_{\rm br}$ for various constant jet luminosities.
Since Eq. \eqref{jet head} overestimates the jet head velocity for $\tilde{L}\lesssim1$ compared with numerical simulations \citep{Mizuta&Ioka2013,Harrison+2018}, we correct Eq. \eqref{jet head} along \cite{Harrison+2018}.
We also modify the collimation condition to $R_{\rm h}\gtrsim\hat{z}/2$ to get a continuous jet cross section.

\begin{figure}
\begin{center}
\includegraphics[width=60mm, angle=90]{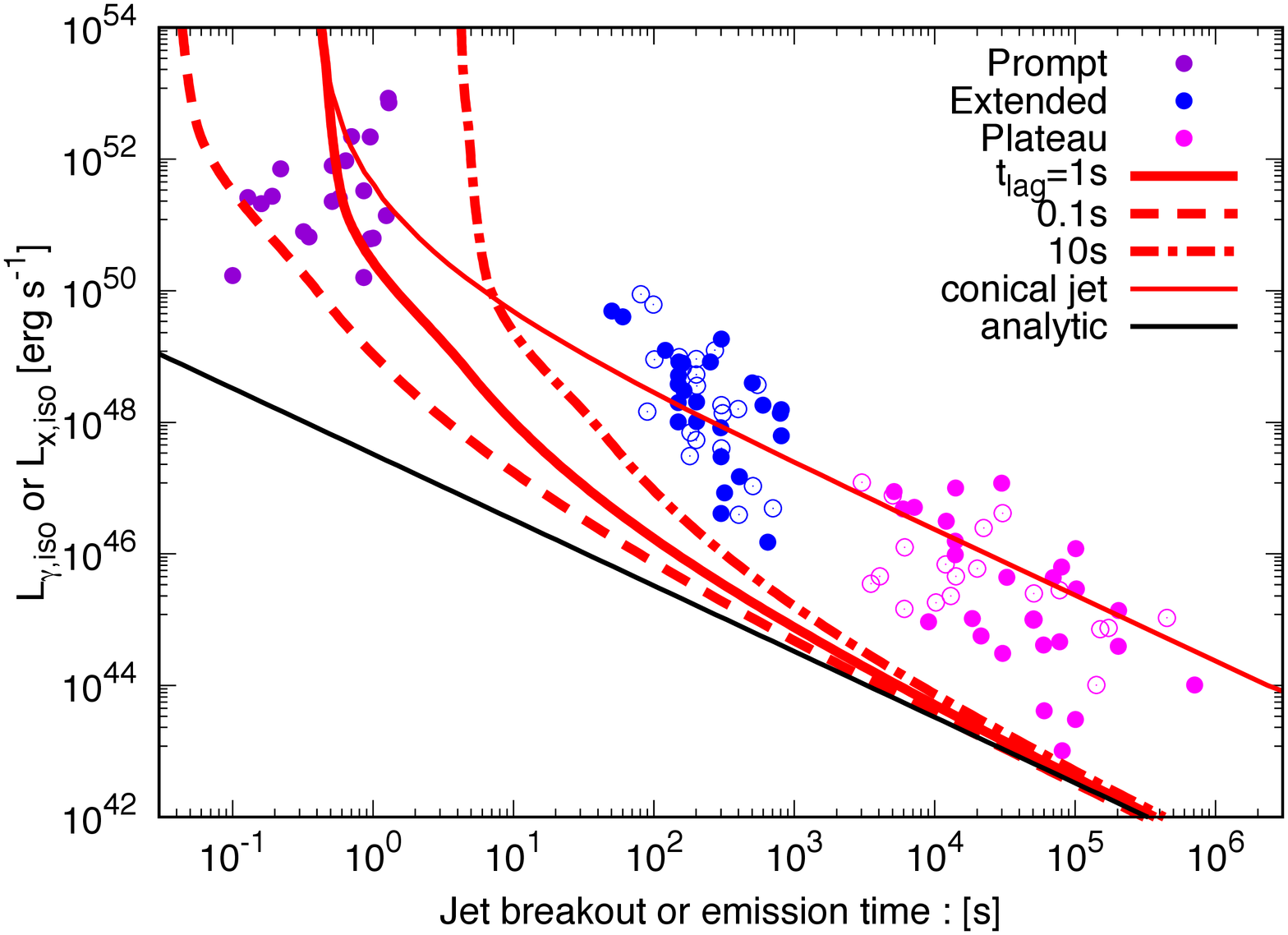}
\caption{
Jet-breakout times for various jet luminosities.
Thick red dashed, solid, and dash-dotted curves show the breakout times for lag-times $t_{\rm lag}=0.1$, $1$, and $10\,\rm s$, respectively.
The other parameters are $\beta_{\rm ej}=0.3$, $\theta_{\rm j}=15^\circ$, $M_{\rm ej}=10^{-2}\,\Msun$, and $k=2$.
The jet and radiation luminosities are related as $L_{\rm \gamma,iso}=\epsilon_{\gamma}{L_{\rm j,iso}}$ and $\epsilon_{\gamma}=0.1$.
Thin red solid curve denotes the result for a conical jet ($t_{\rm lag}=1\,\rm s$).
Black line shows an analytical formula (Eq. \ref{condition matsumoto}).
The data points are taken from \cite{Zou+2018} (for prompt), and \cite{Kisaka+2017} (extended and plateau emissions).
Open circles show the events with unknown redshift (assumed $z=0.72$).}
\label{fig condition}
\end{center}
\end{figure}

In Fig. \ref{fig condition}, we show the result.
Each thick red curve shows the breakout time for each lag-time.
The other parameters are fixed as $\beta_{\rm ej}=0.3$, $\theta_{\rm j}=15^\circ\simeq0.26\,\rm{rad}$, $M_{\rm ej}=10^{-2}\,\Msun$, and $k=2$.
We convert the jet luminosity to the radiation luminosity by adopting an efficiency of $\epsilon_{\gamma}=0.1$ as $L_{\rm \gamma,iso}=\epsilon_{\gamma}L_{\rm j,iso}$.
The ejecta velocity and mass are motivated by numerical simulations \citep{Hotokezaka+2013} and the observations of the macronova in GW170817 \citep[e.g.,][]{Coulter+2017,Utsumi+2017}.
The opening angle is based on the observations of sGRBs \citep{Fong+2015}, while the observed value may be different from the jet-injection angle.
The index $k=2$ is relevant for wind-like ejecta and a larger indices give shorter breakout times.
The thin red curve shows the result for conical jets with $t_{\rm lag}=1\,\rm s$, which give conservative (longer) breakout times.
The emission timescale $t_{\rm em}$ and isotropic luminosity of observed sGRBs' emissions are plotted.
Since the observed sGRBs have successful prompt jets, we can regard the observed emission timescales as engine-working timescales, which ensures that the duration of the engine-activity (jet launching) is long enough for a delayed breakout, $t_{\rm engine}\gtrsim{t_{\rm em}}$.

For a large jet luminosity (e.g., $L_{\rm j,iso}\gtrsim10^{51}{\,\rm erg\,s^{-1}}$ for $t_{\rm lag}=1{\,\rm s}$), the breakout time is smaller than the lag-time $t_{\rm br}\lesssim{t_{\rm lag}}$ and insensitive to the jet luminosity.
This is because a large jet luminosity gives a large jet parameter $\tilde{L}\gtrsim1$ and a jet head velocity becomes almost independent of the jet luminosity $\beta_{\rm h}\sim{1}$.
The breakout time is evaluated by equating the jet head radius $\beta_{\rm h}ct$ and ejecta radius $\beta_{\rm ej}c(t+t_{\rm lag})$ as \citep{Murguia-Berthier+2014},
\begin{eqnarray}
t_{\rm br}\sim\frac{\beta_{\rm ej}t_{\rm lag}}{\beta_{\rm h}-\beta_{\rm ej}}.
   \label{breakout time}
\end{eqnarray}
With $\beta_{\rm h}=1$ and $\beta_{\rm ej}=0.3$, this equation reasonably reproduces our result as $t_{\rm br}\simeq0.4{\,\rm s\,}t_{\rm lag,0}$.
Hereafter, we use the convention $Q_{x}=Q/10^{x}$ (cgs units).
A shorter breakout time than a lag-time enables us to regard that the envelope is static.
In particular, the jet head velocity is constant for the index of $k=2$, which we assumed to derive Eq. \eqref{breakout time}.

For a small jet luminosity, the jet-breakout time gets longer than Eq. \eqref{breakout time} due to a small jet head velocity.
After the lag-time, the expansion of the ejecta affects the jet head dynamics by reducing the ejecta density and accelerating the jet head (see Eq. \ref{L tilde}).
A much longer breakout time than the lag-time is inversely proportional to the jet luminosity $t_{\rm br}\propto{L_{\rm j,iso}^{-1}}$.
Namely, there is a critical energy for a jet to break out of ejecta \citep{Duffell+2018}.
For a conical jet, this energy is simply given by the ejecta energy $M_{\rm ej}(c\beta_{\rm ej})^2/2{\gtrsim}E_{\rm j,iso}\sim{}L_{\rm j,iso}t$, which reasonably reproduces our result $t\gtrsim10^2{\,\rm{s}\,}M_{\rm ej,-2}\beta_{\rm ej,-0.5}^2L_{\rm \gamma,iso,48}^{-1}$.
For a collimated jet with a small ejecta mass in front of the jet head, the required energy is smaller.
In appendix, we derive an analytical scaling law (Eq. \ref{condition matsumoto} and black line in Fig. \ref{fig condition}).
Note that unless the ejecta expansion is taken into account precisely, the breakout time is significantly overestimated except for the parameter dependence \citep[cf.][]{Kimura+2018}.

In particular, a jet-breakout time for a small jet luminosity should be compared with emission timescales of extended ($t_{\rm em}\sim10^{2-3}{\,\rm s}$) and plateau emissions ($t_{\rm em}\sim10^{4-5}{\,\rm s}$).
These emission times are longer than the required breakout time and guarantee that if these emissions are produced by jets, the jets can break out of the ejecta.

\section{Observational Prospects}\label{Delayed Jet Breakout}
We discuss the observational prospects of the delayed breakout events.
In the following, we mainly consider that a late jet producing an extended emission breaks out.
By combining a GW observation and followups, we can check whether a delayed jet breakout occurs or not for a binary merger.
First, such a combination tells us whether the event is on-axis or not \citep{Abbott+2017c,Mandel2018,Finstad+2018}.
For an on-axis event, a detection of prompt $\gamma$-rays tells us the fate of its prompt jet.
If we detect not a prompt emission but an extended (plateau) emission-like signature i.e., a flat light curve up to $\sim10^{2-3}{\,\rm s}$ ($10^{4-5}{\,\rm s}$) and an abrupt shut down, it strongly supports that the late-time jet does punch out a hole in the ejecta.
Therefore, we should threw X-ray detectors to the merger event regardless of whether prompt $\gamma$-rays are detected or not.
In particular, since plateau emissions last for a very long time, they can be a good target  of X-ray detectors such as \textit{Swift} XRT and \textit{MAXI} \citep{Nakamura+2014,Kisaka+2017}.

\subsection{As a probe of late-time engine activity in binary NS mergers}\label{As a probe of late-time engine activity in binary NS mergers}
Delayed-jet-breakout events can be a probe to study what powers extended and plateau emissions.
Currently, the origin of these long-lasting emissions is controversial while there are two representative models.
One is the magnetar model \citep{Metzger+2008,Bucciantini+2012,Rowlinson+2013,Gompertz+2013,Gompertz+2014,Gibson+2017} where a long-lived magnetar powers energetic outflows through the spin-down or propeller effect. 
The outflows dissipate energy and power the emissions.
The other is the black hole (BH) model \citep{Barkov&Pozanenko2011,Nakamura+2014,Kisaka&Ioka2015}, in which the emissions are produced by jets from a BH and accretion disk system fueled by fallback matter \citep{Rosswog2007}. 

Since the delayed jet breakout requires a jet (or a collimated outflow), its detection is an  evidence that the extended or plateau emission is produced by a jet.
Some magnetar models explain the long-lasting emissions by rather isotropic magnetar winds.
The isotropic outflows cannot break out of ejecta by theirselves or produce detectable signals without a hole punched out by a prompt jet.
Therefore, the delayed jet breakout strongly supports a BH jet or a mechanism to collimate isotropic magnetar winds \citep{Bucciantini+2012}.

The jet eventually collides with the interstellar medium (ISM) and produces an afterglow.
The total kinetic energy of the late jet can be comparable to that of prompt jets in ordinary sGRBs.
However, its initial Lorentz factor may be lower than that of normal sGRBs, which causes a different afterglow emission.
Such a jet decelerates at a longer timescale, and its afterglow peaks at $t_{\rm dec}\sim3\times10^{5}\,{\rm s}\,E_{\rm j,iso,51}^{1/3}n_{-4}^{-1/3}\Gamma_1^{-8/3}$, where $\Gamma$ and $n$ are the initial Lorentz factor and the ISM density \citep{Lamb&Kobayashi2016}.
While X-ray and optical afterglows may be dimmer than the following plateau and macronova emissions, an identification of their peaks can be a probe of the Lorentz factor of the late-time jet.

\subsection{Event Rate}
We estimate the event rate of the delayed jet breakout.
The binary-NS-merger rate is evaluated as ${\cal R}_{\rm NSM}\simeq1550_{-1220}^{+3220}{\,\rm Gpc^{-3}\,yr^{-1}}$ by the observation of GW170817 \citep{Abbott+2017c}.
The merger rate for on-axis events is estimated by assuming the jet opening angle to be 

\begin{eqnarray}
{\cal R}_{\rm on}\simeq\frac{\theta_{\rm j}^2}{2}{\cal R}_{\rm NSM}\simeq54_{-42}^{+110}{\,\rm Gpc^{-3}\,yr^{-1}\,}\biggl(\frac{\theta_{\rm j}}{0.26{\,\rm rad}}\biggl)^{2}.
\end{eqnarray}
The central value is larger than the local sGRB rate of $\simeq4.1\,{\rm{Gpc^{-3}\,yr^{-1}}}$ \citep{Wanderman&Piran2015}, and supports that many merger events produce choked jets.
For LIGO's full sensitivity, the detectable range of on-axis binary NS mergers is $d_{\rm L}\simeq1.6\times200{\,\rm Mpc}$, where the factor $1.6$ accounts for an enhancement of GWs \citep{Kochanek&Piran1993}, and the comoving volume is $V_{\rm com}\simeq1.1\times10^{-1}{\,\rm Gpc^3}$.
The on-axis event rate for the observation by LIGO is evaluated as
\begin{eqnarray}
N_{\rm on}=V_{\rm com}{\cal R}_{\rm on}\simeq6.0_{-4.6}^{+12}{\,\rm yr^{-1}\,}.
   \label{on axis rate}
\end{eqnarray}
The fraction of the delayed-jet-breakout events to the total on-axis events (we denote $f_{\rm delay}$) is constrained by the current sky monitors in X- and $\gamma$-rays.
We adopt a luminosity and duration of a late-time jet as $L_{\rm X,iso}=10^{48}{\,\rm erg\,s^{-1}}$ and $t_{\rm em}=300\,\rm s$ as fiducial values.
A detector with a sensitivity $f_{\rm sen}$ can be triggered by the jet inside the luminosity distance of $d_{\rm L}=({L_{\rm X,iso}}/{4\pi{f_{\rm sen}}})^{1/2}\simeq0.94\,{\rm Gpc\,}L_{\rm X,iso,48}^{1/2}f_{\rm sen,-8}^{-1/2}$.
We estimate the detection rate of the extended emissions in delayed jet breakouts by \textit{Swift} BAT and \textit{MAXI} GSC.

BAT has a sensitivity $f_{\rm sen}\sim10^{-8}{\,\rm erg\,s^{-1}\,cm^{-2}}$ and field of view (FoV) $1.4\,\rm str$.
The detection horizon is $d_{\rm{L}}\simeq0.94\,\rm Gpc$ ($V_{\rm com}\simeq2.1{\,\rm Gpc^3}$), and  the sky coverage is $1.4/4\pi\simeq0.11$.
The detection rate is $N_{\rm BAT}\simeq0.1\times0.11\times{V_{\rm com}}f_{\rm delay}{\cal R}_{\rm on}\simeq1.3_{-0.98}^{+2.6}{\,\rm yr^{-1}\,}f_{\rm delay}$,
where we take that BAT has ever detected only $\sim10\,\%$ of extended emissions \citep{Kisaka+2017} into account.
This implies that some soft-long GRBs detected by BAT may be the extended emissions in the delayed jet breakouts.
These events do not accompany supernova (SN) signatures.

\textit{MAXI} has a sensitivity $f_{\rm sen}\sim10^{-9}{\,\rm erg\,s^{-1}\,cm^{-2}}$ for soft bands ($2-30\,\rm keV$) and FoV $7.3\times10^{-2}\,\rm str$ \citep{Sugizaki+2011}.
The horizon and sky coverage are evaluated as $d_{\rm L}\simeq3.0\,\rm Gpc$ ($V_{\rm com}\simeq32{\,\rm Gpc^3}$) and $\simeq5.8\times10^{-3}$, respectively.
The detection rate is $N_{\textit MAXI}\simeq0.0058\times{V_{\rm com}}f_{\rm delay}{\cal R}_{\rm on}\simeq10.0_{-7.8}^{+21}{\,\rm yr^{-1}\,}f_{\rm delay}$.
This value also suggests that some long GRBs detected by \textit{MAXI} are extended emissions in delayed-jet events.
Actually, some GRBs are detected only by \textit{MAXI} and their detection rate is $\sim5\,\rm yr^{-1}$ \citep{Serino+2014}.\footnote{Other X-ray transients might be included in the \textit{MAXI} GRBs, which reduces $f_{\rm delay}$ further.
A candidate is shock breakouts of SNe, while they show thermal emissions.
Detecting the optical counterparts of \textit{MAXI} GRBs, we can firmly distinguish these events from delayed breakouts.}
Therefore, \textit{MAXI} constrains the fraction of delayed events as $f_{\rm delay}\lesssim0.5$ for the central value. Interestingly, for the lower value of ${\cal R}_{\rm NSM}$, the fraction is not constrained due to its small event rate. With the on-axis merger rate (Eq. \ref{on axis rate}), we expect that the rate of delayed jet breakouts coincident with GWs is at most $\sim1-3\,\rm\,yr^{-1}\,(\theta_{\rm j}/0.26\,\rm rad)^2$.\footnote{Note that there is still an uncertainty in the jet opening angle, which affects the event rate. For instance, \cite{Beniamini+2018} argue that the merger rate is consistent with sGRB rate, because they assume a narrower jet-opening angle than ours.}
Future wild-field X-ray monitors such as \textit{ISS-Lobster} \citep{Camp+2013} and \textit{Einstein Probe} \citep{EinsteinProbe2015} will detect delayed events or constrain $f_{\rm delay}$ more tightly.
We also remark that newly-discovered X-ray transients \citep{Bauer+2017} may be related with delayed events.

\section{Discussion}\label{Discussion}
In GW170817, although the VLBI observations revealed a relativistic jet with $E_{\rm j,iso}\gtrsim10^{52}{\,\rm erg}$ \citep{Mooley+2018b,Ghirlanda+2018}, this jet is not necessarily the origin of the low-luminosity prompt $\gamma$-rays \citep{Matsumoto+2018c}.
We can argue that a prompt jet is choked and the resulting cocoon produces the $\gamma$-rays \citep{Kasliwal+2017,Gottlieb+2018b,Lazzati+2018}, and that the relativistic jet is originated from a delayed breakout.
Actually, some extended emissions show $L_{\rm X,iso}\sim10^{49}{\,\rm erg\,s^{-1}}$ with $t_{\rm em}\sim10^{2}\,\rm s$ (Fig. \ref{fig condition}), which suggests a large jet energy of $\sim10^{52}{\,\rm erg}$.
The energetic late jet penetrates the ejecta $\sim10{\,\rm s}$ after the prompt jet is choked, and produces a cocoon with $E_{\rm c}\sim10^{51}{\,\rm erg}$.
Interestingly, this cocoon's cooling emission reproduces the observed macronova at the first few days \citep{Matsumoto+2018b}.
Future GW observations will test this possibility.

In a delayed jet breakout, we have a chance to observe the moment that the late jet to emerge from ejecta.
When the jet head reaches the ejecta edge, it may produce a shock-breakout emission \citep{Gottlieb+2018b}.
Even if a merger event occurs outside the FoV of $\gamma$-ray telescopes, the detection of this breakout signature may support the delayed jet breakout.

The delayed-breakout events can be a source of neutrinos.
A choked prompt jet can be a powerful neutrino emitter \citep{Kimura+2018}.
Moreover, a delayed jet emits neutrinos efficiently if it has a low Lorentz factor \citep{Kimura+2017}.
Detections of these neutrinos can constrain the Lorentz factors and the baryon loading of these jets. 

\cite{Lamb&Kobayashi2016} propose another scenario where on-axis binary NS mergers do not produce $\gamma$-rays.
They consider that a low-Lorentz-factor prompt jet breaks out of ejecta but does not emit $\gamma$-rays due to the compactness, and discuss the detectability of its afterglow.
On the other hand, our scenario predicts that an extended or plateau emission accompanies with the merger.
In particular, a flat light curve and a sudden drop are unique signatures of a central engine activity.

Finally, we discuss the breakout condition of prompt jets.
A comparison of the breakout time (red curves in Fig. \ref{fig condition}) with the prompt-emission timescale (purple points) suggests that the lag-time should be smaller than $t_{\rm lag}\lesssim1{\,\rm s}$ to produce a sGRB \citep[see also][]{Murguia-Berthier+2014}.
For longer lag-times, a jet cannot catch up with the ejecta edge within the engine-working time.
{Note that an event with a short emission time ($t_{\rm em}\ll t_{\rm br}$) does not constrain the lag-time because such a burst is produced by a bare breakout $t_{\rm engine}=t_{\rm br}+t_{\rm em}\sim{t_{\rm br}}$}

If each merger event has a similar ejecta velocity and a common lag-time, a characteristic breakout time is introduced (see Eq. \ref{breakout time}).
Intriguingly, \cite{Moharana&Piran2017} found a typical jet-breakout time $t_{\rm br}\simeq0.2-0.5{\,\rm s}$.
In the collapsar scenario, such a timescale is understood as a time for a jet to reach the progenitor's edge whose size may not change significantly among progenitors \citep{Bromberg+2012}.
However, binary mergers do not have a characteristic size because ejecta expand.
A common lag-time introduces such a special length $\sim\beta_{\rm ej}ct_{\rm lag}$ into the systems.
Therefore, the typical breakout time may suggest that there is a favored lag-time to produce sGRBs, which may be related to the jet-launch mechanism such as formations of global magnetic fields or a BH.

\acknowledgments%\section*{acknowledgments}
We thank Motoko Serino for useful comments on the observations by \textit{MAXI}.
We are also grateful to Peter M\'{e}sz\'{a}ros for helpful comments.
This work is supported by JSPS Overseas Challenge Program for Young Researchers, Grant-in-Aid for JSPS Research Fellow 17J09895 (T.M.), JSPS Oversea Research Fellowship, and the IGC post-doctoral fellowship program (S.S.K.).

%%%%%%%%%%%%%%%%%%%%%%%%%%%%%%%%%%%%%%%%%%%%%%%%%%

\appendix
\section{Analytical Formula}\label{appendix}
We derive analytical formulae for jet propagations in homologously expanding media.

\subsection{Conical jet}
For a conical jet, the jet parameter is given by
\begin{eqnarray}
{\tilde L}=\frac{L_{\rm j}}{\pi\theta_{\rm j}^2R_{\rm h}^2\rho_{\rm a}c^3}=\frac{L_{\rm j}}{\pi{\cal A}\theta_{\rm j}^2c^3}R_{\rm h}^{k-2},
\end{eqnarray}
where we define $\rho_{\rm a}={\cal A}R_{\rm h}^{-k}$.
For $\tilde L\lesssim1$, we can approximate Eq. \eqref{jet head} as $\beta_{\rm h}\simeq\tilde{L}^{1/2}+\beta_{\rm a}$ and rewrite it as
\begin{eqnarray}
t\frac{d(R_{\rm h}/t)}{dt}=c\tilde{L}^{1/2},
   \label{differential equation}
\end{eqnarray}
where we used $\beta_{\rm a}\simeq{R_{\rm h}}/t$ for a homologously expanding media at later than $t>t_{\rm lag}$.
Due to a term $\beta_{\rm a}$, the left hand side has a different form than that for static media cases.
This gives a different numerical coefficient than static one.
By integrating Eq. \eqref{differential equation}, we obtain a formula 
\begin{eqnarray}
R_{\rm h}=\biggl[\frac{L_{\rm j}}{{\cal A}\theta_{\rm j}^2c}\frac{1}{\pi}\biggl(\frac{4-k}{k-p-2}\biggl)^2\biggl]^{\frac{1}{4-k}}t^{\frac{2}{4-k}}\propto{t^{\frac{2-p}{4-k}}},
   \label{jet head position ana}
\end{eqnarray}
for $2+p<k<4$, where the constant $p$ is defined as ${\cal A}=\tilde{\cal A}t^p$.
In our model, the quantities are $k=2$, ${\cal A}=M_{\rm ej}f/4\pi{R_{\rm ej}}^{3-k}$, and $p=k-3$, and give
\begin{eqnarray}
R_{\rm h}=7.6\times10^{6}{\,\rm cm\,}M_{\rm ej,-2}^{-1/2}L_{\rm j,iso,45}^{1/2}\beta_{\rm ej,-0.5}^{1/2}(t/{\rm s})^{3/2}.
\end{eqnarray}
By equating this radius with the ejecta edge $R_{\rm ej}\sim\beta_{\rm ej}ct$, we obtain the critical luminosity for successful jet breakouts as
\begin{eqnarray}
L_{\rm j,iso}\simeq1.5\times10^{51}{\,\rm erg\,s^{-1}\,}M_{\rm ej,-2}\beta_{\rm ej,-0.5}(t/{\rm s})^{-1}.
\end{eqnarray}

\subsection{Collimated jet}
Next, we consider a collimated jet.
At $t>t_{\rm lag}$, we denote the cocoon radus as $R_{\rm c}=\xi\beta_{\rm c}ct$, where a numerical coefficient $\xi$ is given later.
The cocoon pressure is rewritten by using Eqs. \eqref{cocoon pressure} and \eqref{cocoon velocity} as
\begin{eqnarray}
P_{\rm c}\simeq\frac{L_{\rm j}t}{3\frac{\pi}{3}R_{\rm c}^2R_{\rm h}}=\frac{L_{\rm j}}{\pi\xi^2\beta_{\rm c}^2R_{\rm h}c^2t}=\biggl(\frac{L_{\rm j}\bar{\rho}_{\rm a}}{\pi\xi^2R_{\rm h}t}\biggl)^{1/2}.
\end{eqnarray}
The jet parameter is given as
\begin{eqnarray}
\tilde{L}=\frac{L_{\rm j}}{\frac{L_{\rm j}\theta_{\rm j}^2}{4cP_{\rm c}}\rho_{\rm a}c^3}=\biggl(\frac{L_{\rm j}}{{\cal A}\theta_{\rm j}^4c^4}\frac{16\varrho}{\pi\xi^2}\frac{R_{\rm h}^{k-1}}{t}\biggl)^{1/2},
\end{eqnarray}
where we use $\bar{\rho}_{\rm a}=\varrho\rho_{\rm a}$ and $\varrho=3/(3-k)$ for a conical cocoon.
By substituting this for Eq. \eqref{differential equation}, we obtain
\begin{eqnarray}
R_{\rm h}=\biggl[\frac{L_{\rm j}}{{\cal A}\theta_{\rm j}^4}\frac{16\varrho}{\pi\xi^2}\biggl(\frac{5-k}{k-2-p}\biggl)^4\biggl]^{\frac{1}{5-k}}t^{\frac{3}{5-k}}\propto{t}^{\frac{3-p}{5-k}},
\end{eqnarray}
for $p-2<k<5$.
The cocoon velocity is given as
\begin{eqnarray}
\beta_{\rm c}\propto(P_{\rm c}/\bar{\rho}_{\rm a})^{1/2}\propto(\rho_{\rm a}R_{\rm h}t)^{-1/4}\propto(t^{p+1}R_{\rm h}^{1-k})^{-1/4}\propto{t^{\frac{p+2-k}{k-5}}},
\end{eqnarray}
which gives the coefficient as $\xi=(5-k)/(3-p)$.
Finally, we get analytical expressions as
\begin{eqnarray}
R_{\rm h}&=&N_{\rm s}^{\frac{5}{5-k}}\biggl[\frac{3L_{\rm j}}{{\cal A}\theta_{\rm j}^4}\frac{2^4}{\pi(3-p)^2}\frac{(5-k)^{2}}{(3-k)}\biggl(\frac{3-p}{p+2-k}\biggl)^4\biggl]^{\frac{1}{5-k}}t^{\frac{3}{5-k}},\\
\beta_{\rm h}&=&N_{\rm s}^{\frac{5}{5-k}}\biggl[\frac{3L_{\rm j}}{{\cal A}\theta_{\rm j}^4}\frac{2^4(3-p)^{3-k}}{\pi}\frac{(5-k)^{k-3}}{(3-k)}\biggl(\frac{3-p}{p+2-k}\biggl)^{4}\biggl]^{\frac{1}{5-k}}\frac{t^{\frac{k-2}{5-k}}}{c},
\end{eqnarray}
where we introduce a correction factor $N_{\rm s}(=0.35\,\text{for\,}\tilde{L}<1)$ given by \cite{Harrison+2018}.
The time dependences are the same as \cite{Margalit+2018}.
We compare these forms with Eqs. (A2) and (A3) in \cite{Harrison+2018}.
They do not consider the time-dependent $\cal A$, which modifies numerical coefficients.
Furthermore, in expanding media, the jet head velocity is determined by not only $\tilde L$ but also $\beta_{\rm a}$ (see Eq. \ref{jet head}).
In particular, Eq. \eqref{differential equation} gives another factor $[(3-p)/(p+2-k)]^4$.

For our case, the jet head velocity and radius are given by
\begin{eqnarray}
\beta_{\rm h}&=&3.0\times10^{-2}\,M_{\rm ej,-2}^{-1/3}L_{\rm j,iso,45}^{1/3}\beta_{\rm ej,-0.5}^{1/3}\theta_{\rm ej,-0.5}^{-2/3}(t/{\rm s})^{1/3}\biggl(\frac{N_{\rm s}}{0.35}\biggl)^{5/3},\\
R_{\rm h}&=&6.8\times10^{8}{\,\rm cm\,}M_{\rm ej,-2}^{-1/3}L_{\rm j,iso,45}^{1/3}\beta_{\rm ej,-0.5}^{1/3}\theta_{\rm ej,-0.5}^{-2/3}(t/{\rm s})^{4/3}\biggl(\frac{N_{\rm s}}{0.35}\biggl)^{5/3},
\end{eqnarray}
and the critical luminosity is given by
\begin{eqnarray}
L_{\rm j,iso}=5.4\times10^{48}{\,\rm erg\,s^{-1}\,}M_{\rm ej,-2}\beta_{\rm ej,-0.5}^{2}\theta_{\rm ej,-0.5}^{2}(t/{\rm s})^{-1}\biggl(\frac{N_{\rm s}}{0.35}\biggl)^{-5}.
   \label{condition matsumoto}
\end{eqnarray}
It should be noted that the different numerical factors introduce a large difference in the critical luminosity.
Actually \cite{Kimura+2018} use the equations in \cite{Harrison+2018} and obtained a much larger critical luminosity than Eq. \eqref{condition matsumoto}.
The discrepancy between ours and theirs are reasonably attributed to the different numerical factor which they adopted as $[(3-p)/3]^{2}[(p+2-k)/(3-p)]^4\simeq7\times10^{-3}$.

% Don't change these lines
% typesetting comment
%\bsp
%\label{lastpage}

\bibliographystyle{apj}
\bibliography{apj-jour,reference_matsumoto}

\begin{thebibliography}{}
\expandafter\ifx\csname natexlab\endcsname\relax\def\natexlab#1{#1}\fi

\bibitem[{{Abbott} {et~al.}(2017){Abbott}, {Abbott}, {Abbott}, {Acernese},
  {Ackley}, {Adams}, {Adams}, {Addesso}, {Adhikari}, {Adya}, \&
  et~al.}]{Abbott+2017c}
{Abbott}, B.~P., {Abbott}, R., {Abbott}, T.~D., {et~al.} 2017, Physical Review
  Letters, 119, 161101

\bibitem[{{Barkov} \& {Pozanenko}(2011)}]{Barkov&Pozanenko2011}
{Barkov}, M.~V., \& {Pozanenko}, A.~S. 2011, \mnras, 417, 2161

\bibitem[{{Bauer} {et~al.}(2017){Bauer}, {Treister}, {Schawinski}, {Schulze},
  {Luo}, {Alexander}, {Brandt}, {Comastri}, {Forster}, {Gilli}, {Kann},
  {Maeda}, {Nomoto}, {Paolillo}, {Ranalli}, {Schneider}, {Shemmer}, {Tanaka},
  {Tolstov}, {Tominaga}, {Tozzi}, {Vignali}, {Wang}, {Xue}, \&
  {Yang}}]{Bauer+2017}
{Bauer}, F.~E., {Treister}, E., {Schawinski}, K., {et~al.} 2017, \mnras, 467,
  4841

\bibitem[{{Begelman} \& {Cioffi}(1989)}]{Begelman&Cioffi1989}
{Begelman}, M.~C., \& {Cioffi}, D.~F. 1989, \apjl, 345, L21

\bibitem[{{Beniamini} {et~al.}(2018){Beniamini}, {Petropoulou}, {Barniol
  Duran}, \& {Giannios}}]{Beniamini+2018}
{Beniamini}, P., {Petropoulou}, M., {Barniol Duran}, R., \& {Giannios}, D.
  2018, ArXiv e-prints, arXiv:1808.04831

\bibitem[{{Berger}(2014)}]{Berger2014}
{Berger}, E. 2014, \araa, 52, 43

\bibitem[{{Bromberg} {et~al.}(2011){Bromberg}, {Nakar}, {Piran}, \&
  {Sari}}]{Bromberg+2011}
{Bromberg}, O., {Nakar}, E., {Piran}, T., \& {Sari}, R. 2011, \apj, 740, 100

\bibitem[{{Bromberg} {et~al.}(2012){Bromberg}, {Nakar}, {Piran}, \&
  {Sari}}]{Bromberg+2012}
---. 2012, \apj, 749, 110

\bibitem[{{Bucciantini} {et~al.}(2012){Bucciantini}, {Metzger}, {Thompson}, \&
  {Quataert}}]{Bucciantini+2012}
{Bucciantini}, N., {Metzger}, B.~D., {Thompson}, T.~A., \& {Quataert}, E. 2012,
  \mnras, 419, 1537

\bibitem[{{Camp} {et~al.}(2013){Camp}, {Barthelmy}, {Blackburn}, {Carpenter},
  {Gehrels}, {Kanner}, {Marshall}, {Racusin}, \& {Sakamoto}}]{Camp+2013}
{Camp}, J., {Barthelmy}, S., {Blackburn}, L., {et~al.} 2013, Experimental
  Astronomy, 36, 505

\bibitem[{{Coulter} {et~al.}(2017){Coulter}, {Foley}, {Kilpatrick}, {Drout},
  {Piro}, {Shappee}, {Siebert}, {Simon}, {Ulloa}, {Kasen}, {Madore},
  {Murguia-Berthier}, {Pan}, {Prochaska}, {Ramirez-Ruiz}, {Rest}, \&
  {Rojas-Bravo}}]{Coulter+2017}
{Coulter}, D.~A., {Foley}, R.~J., {Kilpatrick}, C.~D., {et~al.} 2017, Science,
  358, 1556

\bibitem[{{Duffell} {et~al.}(2018){Duffell}, {Quataert}, {Kasen}, \&
  {Klion}}]{Duffell+2018}
{Duffell}, P.~C., {Quataert}, E., {Kasen}, D., \& {Klion}, H. 2018, ArXiv
  e-prints, arXiv:1806.10616

\bibitem[{{Eichler} {et~al.}(1989){Eichler}, {Livio}, {Piran}, \&
  {Schramm}}]{Eichler+1989}
{Eichler}, D., {Livio}, M., {Piran}, T., \& {Schramm}, D.~N. 1989, \nat, 340,
  126

\bibitem[{{Finstad} {et~al.}(2018){Finstad}, {De}, {Brown}, {Berger}, \&
  {Biwer}}]{Finstad+2018}
{Finstad}, D., {De}, S., {Brown}, D.~A., {Berger}, E., \& {Biwer}, C.~M. 2018,
  \apjl, 860, L2

\bibitem[{{Fong} {et~al.}(2015){Fong}, {Berger}, {Margutti}, \&
  {Zauderer}}]{Fong+2015}
{Fong}, W., {Berger}, E., {Margutti}, R., \& {Zauderer}, B.~A. 2015, \apj, 815,
  102

\bibitem[{{Ghirlanda} {et~al.}(2018){Ghirlanda}, {Salafia}, {Paragi},
  {Giroletti}, {Yang}, {Marcote}, {Blanchard}, {Agudo}, {An}, {Bernardini},
  {Beswick}, {Branchesi}, {Campana}, {Casadio}, {Chassande-Mottin}, {Colpi},
  {Covino}, {D'Avanzo}, {D'Elia}, {Frey}, {Gawronski}, {Ghisellini}, {Gurvits},
  {Jonker}, {van Langevelde}, {Melandri}, {Moldon}, {Nava}, {Perego},
  {Perez-Torres}, {Reynolds}, {Salvaterra}, {Tagliaferri}, {Venturi},
  {Vergani}, \& {Zhang}}]{Ghirlanda+2018}
{Ghirlanda}, G., {Salafia}, O.~S., {Paragi}, Z., {et~al.} 2018, ArXiv e-prints,
  arXiv:1808.00469

\bibitem[{{Gibson} {et~al.}(2017){Gibson}, {Wynn}, {Gompertz}, \&
  {O'Brien}}]{Gibson+2017}
{Gibson}, S.~L., {Wynn}, G.~A., {Gompertz}, B.~P., \& {O'Brien}, P.~T. 2017,
  \mnras, 470, 4925

\bibitem[{{Gompertz} {et~al.}(2014){Gompertz}, {O'Brien}, \&
  {Wynn}}]{Gompertz+2014}
{Gompertz}, B.~P., {O'Brien}, P.~T., \& {Wynn}, G.~A. 2014, \mnras, 438, 240

\bibitem[{{Gompertz} {et~al.}(2013){Gompertz}, {O'Brien}, {Wynn}, \&
  {Rowlinson}}]{Gompertz+2013}
{Gompertz}, B.~P., {O'Brien}, P.~T., {Wynn}, G.~A., \& {Rowlinson}, A. 2013,
  \mnras, 431, 1745

\bibitem[{{Gottlieb} {et~al.}(2018){Gottlieb}, {Nakar}, {Piran}, \&
  {Hotokezaka}}]{Gottlieb+2018b}
{Gottlieb}, O., {Nakar}, E., {Piran}, T., \& {Hotokezaka}, K. 2018, \mnras,
  arXiv:1710.05896

\bibitem[{{Harrison} {et~al.}(2018){Harrison}, {Gottlieb}, \&
  {Nakar}}]{Harrison+2018}
{Harrison}, R., {Gottlieb}, O., \& {Nakar}, E. 2018, \mnras, 477, 2128

\bibitem[{{Hotokezaka} {et~al.}(2013){Hotokezaka}, {Kiuchi}, {Kyutoku},
  {Okawa}, {Sekiguchi}, {Shibata}, \& {Taniguchi}}]{Hotokezaka+2013}
{Hotokezaka}, K., {Kiuchi}, K., {Kyutoku}, K., {et~al.} 2013, \prd, 87, 024001

\bibitem[{{Ioka} {et~al.}(2005){Ioka}, {Kobayashi}, \& {Zhang}}]{Ioka+2005}
{Ioka}, K., {Kobayashi}, S., \& {Zhang}, B. 2005, \apj, 631, 429

\bibitem[{{Kasliwal} {et~al.}(2017){Kasliwal}, {Nakar}, {Singer}, {Kaplan},
  {Cook}, {Van Sistine}, {Lau}, {Fremling}, {Gottlieb}, {Jencson}, {Adams},
  {Feindt}, {Hotokezaka}, {Ghosh}, {Perley}, {Yu}, {Piran}, {Allison},
  {Anupama}, {Balasubramanian}, {Bannister}, {Bally}, {Barnes}, {Barway},
  {Bellm}, {Bhalerao}, {Bhattacharya}, {Blagorodnova}, {Bloom}, {Brady},
  {Cannella}, {Chatterjee}, {Cenko}, {Cobb}, {Copperwheat}, {Corsi}, {De},
  {Dobie}, {Emery}, {Evans}, {Fox}, {Frail}, {Frohmaier}, {Goobar}, {Hallinan},
  {Harrison}, {Helou}, {Hinderer}, {Ho}, {Horesh}, {Ip}, {Itoh}, {Kasen},
  {Kim}, {Kuin}, {Kupfer}, {Lynch}, {Madsen}, {Mazzali}, {Miller}, {Mooley},
  {Murphy}, {Ngeow}, {Nichols}, {Nissanke}, {Nugent}, {Ofek}, {Qi}, {Quimby},
  {Rosswog}, {Rusu}, {Sadler}, {Schmidt}, {Sollerman}, {Steele}, {Williamson},
  {Xu}, {Yan}, {Yatsu}, {Zhang}, \& {Zhao}}]{Kasliwal+2017}
{Kasliwal}, M.~M., {Nakar}, E., {Singer}, L.~P., {et~al.} 2017, Science, 358,
  1559

\bibitem[{{Kimura} {et~al.}(2018){Kimura}, {Murase}, {Bartos}, {Ioka}, {Heng},
  \& {M{\'e}sz{\'a}ros}}]{Kimura+2018}
{Kimura}, S.~S., {Murase}, K., {Bartos}, I., {et~al.} 2018, ArXiv e-prints,
  arXiv:1805.11613

\bibitem[{{Kimura} {et~al.}(2017){Kimura}, {Murase}, {M{\'e}sz{\'a}ros}, \&
  {Kiuchi}}]{Kimura+2017}
{Kimura}, S.~S., {Murase}, K., {M{\'e}sz{\'a}ros}, P., \& {Kiuchi}, K. 2017,
  \apjl, 848, L4

\bibitem[{{Kisaka} \& {Ioka}(2015)}]{Kisaka&Ioka2015}
{Kisaka}, S., \& {Ioka}, K. 2015, \apjl, 804, L16

\bibitem[{{Kisaka} {et~al.}(2017){Kisaka}, {Ioka}, \& {Sakamoto}}]{Kisaka+2017}
{Kisaka}, S., {Ioka}, K., \& {Sakamoto}, T. 2017, \apj, 846, 142

\bibitem[{{Kochanek} \& {Piran}(1993)}]{Kochanek&Piran1993}
{Kochanek}, C.~S., \& {Piran}, T. 1993, \apjl, 417, L17

\bibitem[{{Komissarov} \& {Falle}(1997)}]{Komissarov&Falle1997}
{Komissarov}, S.~S., \& {Falle}, S.~A.~E.~G. 1997, \mnras, 288, 833

\bibitem[{{Lamb} \& {Kobayashi}(2016)}]{Lamb&Kobayashi2016}
{Lamb}, G.~P., \& {Kobayashi}, S. 2016, \apj, 829, 112

\bibitem[{{Lazzati} {et~al.}(2018){Lazzati}, {Perna}, {Morsony},
  {Lopez-Camara}, {Cantiello}, {Ciolfi}, {Giacomazzo}, \&
  {Workman}}]{Lazzati+2018}
{Lazzati}, D., {Perna}, R., {Morsony}, B.~J., {et~al.} 2018, Physical Review
  Letters, 120, 241103

\bibitem[{{Mandel}(2018)}]{Mandel2018}
{Mandel}, I. 2018, \apjl, 853, L12

\bibitem[{{Margalit} {et~al.}(2018){Margalit}, {Metzger}, {Thompson},
  {Nicholl}, \& {Sukhbold}}]{Margalit+2018}
{Margalit}, B., {Metzger}, B.~D., {Thompson}, T.~A., {Nicholl}, M., \&
  {Sukhbold}, T. 2018, \mnras, 475, 2659

\bibitem[{{Matsumoto} {et~al.}(2018{\natexlab{a}}){Matsumoto}, {Ioka},
  {Kisaka}, \& {Nakar}}]{Matsumoto+2018b}
{Matsumoto}, T., {Ioka}, K., {Kisaka}, S., \& {Nakar}, E. 2018{\natexlab{a}},
  \apj, 861, 55

\bibitem[{{Matsumoto} {et~al.}(2018{\natexlab{b}}){Matsumoto}, {Nakar}, \&
  {Piran}}]{Matsumoto+2018c}
{Matsumoto}, T., {Nakar}, E., \& {Piran}, T. 2018{\natexlab{b}}, ArXiv
  e-prints, arXiv:1807.04756

\bibitem[{{Matzner}(2003)}]{Matzner2003}
{Matzner}, C.~D. 2003, \mnras, 345, 575

\bibitem[{{Metzger} {et~al.}(2008){Metzger}, {Quataert}, \&
  {Thompson}}]{Metzger+2008}
{Metzger}, B.~D., {Quataert}, E., \& {Thompson}, T.~A. 2008, \mnras, 385, 1455

\bibitem[{{Mizuta} \& {Ioka}(2013)}]{Mizuta&Ioka2013}
{Mizuta}, A., \& {Ioka}, K. 2013, \apj, 777, 162

\bibitem[{{Moharana} \& {Piran}(2017)}]{Moharana&Piran2017}
{Moharana}, R., \& {Piran}, T. 2017, \mnras, 472, L55

\bibitem[{{Mooley} {et~al.}(2018){Mooley}, {Deller}, {Gottlieb}, {Nakar},
  {Hallinan}, {Bourke}, {Frail}, {Horesh}, {Corsi}, \&
  {Hotokezaka}}]{Mooley+2018b}
{Mooley}, K.~P., {Deller}, A.~T., {Gottlieb}, O., {et~al.} 2018, ArXiv
  e-prints, arXiv:1806.09693

\bibitem[{{Murguia-Berthier} {et~al.}(2014){Murguia-Berthier}, {Montes},
  {Ramirez-Ruiz}, {De Colle}, \& {Lee}}]{Murguia-Berthier+2014}
{Murguia-Berthier}, A., {Montes}, G., {Ramirez-Ruiz}, E., {De Colle}, F., \&
  {Lee}, W.~H. 2014, \apjl, 788, L8

\bibitem[{{Nagakura} {et~al.}(2014){Nagakura}, {Hotokezaka}, {Sekiguchi},
  {Shibata}, \& {Ioka}}]{Nagakura+2014}
{Nagakura}, H., {Hotokezaka}, K., {Sekiguchi}, Y., {Shibata}, M., \& {Ioka}, K.
  2014, \apjl, 784, L28

\bibitem[{{Nakamura} {et~al.}(2014){Nakamura}, {Kashiyama}, {Nakauchi}, {Suwa},
  {Sakamoto}, \& {Kawai}}]{Nakamura+2014}
{Nakamura}, T., {Kashiyama}, K., {Nakauchi}, D., {et~al.} 2014, \apj, 796, 13

\bibitem[{{Nakar}(2007)}]{Nakar2007}
{Nakar}, E. 2007, \physrep, 442, 166

\bibitem[{{Norris} \& {Bonnell}(2006)}]{Norris&Bonnell2006}
{Norris}, J.~P., \& {Bonnell}, J.~T. 2006, \apj, 643, 266

\bibitem[{{Perley} {et~al.}(2009){Perley}, {Metzger}, {Granot}, {Butler},
  {Sakamoto}, {Ramirez-Ruiz}, {Levan}, {Bloom}, {Miller}, {Bunker}, {Chen},
  {Filippenko}, {Gehrels}, {Glazebrook}, {Hall}, {Hurley}, {Kocevski}, {Li},
  {Lopez}, {Norris}, {Piro}, {Poznanski}, {Prochaska}, {Quataert}, \&
  {Tanvir}}]{Perley+2009}
{Perley}, D.~A., {Metzger}, B.~D., {Granot}, J., {et~al.} 2009, \apj, 696, 1871

\bibitem[{{Rosswog}(2007)}]{Rosswog2007}
{Rosswog}, S. 2007, \mnras, 376, L48

\bibitem[{{Rowlinson} {et~al.}(2013){Rowlinson}, {O'Brien}, {Metzger},
  {Tanvir}, \& {Levan}}]{Rowlinson+2013}
{Rowlinson}, A., {O'Brien}, P.~T., {Metzger}, B.~D., {Tanvir}, N.~R., \&
  {Levan}, A.~J. 2013, \mnras, 430, 1061

\bibitem[{{Serino} {et~al.}(2014){Serino}, {Sakamoto}, {Kawai}, {Yoshida},
  {Ohno}, {Ogawa}, {Nishimura}, {Fukushima}, {Higa}, {Ishikawa}, {Ishikawa},
  {Kawamuro}, {Kimura}, {Matsuoka}, {Mihara}, {Morii}, {Nakagawa}, {Nakahira},
  {Nakajima}, {Nakano}, {Negoro}, {Onodera}, {Sasaki}, {Shidatsu}, {Sugimoto},
  {Sugizaki}, {Suwa}, {Suzuki}, {Tachibana}, {Takagi}, {Toizumi}, {Tomida},
  {Tsuboi}, {Tsunemi}, {Ueda}, {Ueno}, {Usui}, {Yamada}, {Yamamoto}, {Yamaoka},
  {Yamauchi}, {Yoshidome}, \& {Yoshii}}]{Serino+2014}
{Serino}, M., {Sakamoto}, T., {Kawai}, N., {et~al.} 2014, \pasj, 66, 87

\bibitem[{{Sugizaki} {et~al.}(2011){Sugizaki}, {Mihara}, {Serino}, {Yamamoto},
  {Matsuoka}, {Kohama}, {Tomida}, {Ueno}, {Kawai}, {Morii}, {Sugimori},
  {Nakahira}, {Yamaoka}, {Yoshida}, {Nakajima}, {Negoro}, {Eguchi}, {Isobe},
  {Ueda}, \& {Tsunemi}}]{Sugizaki+2011}
{Sugizaki}, M., {Mihara}, T., {Serino}, M., {et~al.} 2011, \pasj, 63, S635

\bibitem[{{Utsumi} {et~al.}(2017){Utsumi}, {Tanaka}, {Tominaga}, {Yoshida},
  {Barway}, {Nagayama}, {Zenko}, {Aoki}, {Fujiyoshi}, {Furusawa}, {Kawabata},
  {Koshida}, {Lee}, {Morokuma}, {Motohara}, {Nakata}, {Ohsawa}, {Ohta},
  {Okita}, {Tajitsu}, {Tanaka}, {Terai}, {Yasuda}, {Abe}, {Asakura}, {Bond},
  {Miyazaki}, {Sumi}, {Tristram}, {Honda}, {Itoh}, {Itoh}, {Kawabata},
  {Morihana}, {Nagashima}, {Nakaoka}, {Ohshima}, {Takahashi}, {Takayama},
  {Aoki}, {Baar}, {Doi}, {Finet}, {Kanda}, {Kawai}, {Kim}, {Kuroda}, {Liu},
  {Matsubayashi}, {Murata}, {Nagai}, {Saito}, {Saito}, {Sako}, {Sekiguchi},
  {Tamura}, {Tanaka}, {Uemura}, \& {Yamaguchi}}]{Utsumi+2017}
{Utsumi}, Y., {Tanaka}, M., {Tominaga}, N., {et~al.} 2017, \pasj, 69, 101

\bibitem[{{Wanderman} \& {Piran}(2015)}]{Wanderman&Piran2015}
{Wanderman}, D., \& {Piran}, T. 2015, \mnras, 448, 3026

\bibitem[{{Yuan} {et~al.}(2015){Yuan}, {Zhang}, {Feng}, {Zhang}, {Ling},
  {Zhao}, {Deng}, {Qiu}, {Osborne}, {O'Brien}, {Willingale}, {Lapington},
  {Fraser}, \& {the Einstein Probe team}}]{EinsteinProbe2015}
{Yuan}, W., {Zhang}, C., {Feng}, H., {et~al.} 2015, ArXiv e-prints,
  arXiv:1506.07735

\bibitem[{{Zou} {et~al.}(2018){Zou}, {Wang}, {Moharana}, {Liao}, {Chen}, {Wu},
  {Lei}, \& {Wang}}]{Zou+2018}
{Zou}, Y.-C., {Wang}, F.-F., {Moharana}, R., {et~al.} 2018, \apjl, 852, L1

\end{thebibliography}

\if{

}\fi

\end{document}